# Full polarization control for fiber optical quantum communication systems using polarization encoding

G. B. Xavier, G. Vilela de Faria, G. P. Temporão and J. P. von der Weid*

*Pontifical Catholic University of Rio de Janeiro - Optoelectronics & Instrumentation Group*
*Center for Telecommunications Studies – Rio de Janeiro - Brazil*
*\*Corresponding author: vdweid@cpti.cetuc.puc-rio.br*

**Abstract:** A real-time polarization control system employing two non-orthogonal reference signals multiplexed in either time or wavelength with the data signal is presented. It is shown, theoretically and experimentally, that complete control of multiple polarization states can be attained employing polarization controllers in closed-loop configuration. Experimental results show that negligible added penalties, corresponding to an average added optical Quantum Bit Error Rate of 0.044%, can be achieved with response times smaller than 10 ms, without significant introduction of noise counts in the quantum channel.

## 1. Introduction

Polarization Mode Dispersion (PMD) is a well known problem in optical transmission systems [1]. It is generated by random fluctuations of the residual birefringence in optical fibers, such that the State of Polarization (SOP) of an optical signal will change randomly over time, in an unpredictable way. If one wishes to keep a correlation between the SOPs at the input and output of a transmission link, some type of active polarization stabilization is needed [2]. A typical example of this situation is found in coherent transmissions, where the polarization state of the received signal must match a fixed SOP of the local oscillator for maximum interference. A more complicated situation is found when different independent polarization states must be mapped into corresponding states at the receiver; this is precisely the case of fiber-optical Quantum Communication systems employing polarization coding, where a quantum bit (qubit) is assigned to the SOP of the transmitted single photon [3]. In this case, the sender of the message randomly chooses one out of two independent bases which are mutually non-orthogonal to code its qubits and the receiver must read the data using a similar procedure. Clearly, any random polarization rotations caused to the signal would make transmission of quantum states impossible unless a full polarization control, able to simultaneously control multiple polarization states, can be achieved.

In this most general case, every input and output pair of polarization states are required to be kept identical (or related by a fixed unitary transformation) over time. The role of such a polarization control system is to insert a unitary transformation immediately before the receiver such that the Jones matrix of the global system (fiber + controller) is the identity. In this letter, we present a scheme for polarization control that employs two non-orthogonal reference signals, multiplexed either in time or wavelength with the signal to be controlled. Each reference is used to generate a feedback signal which is used to perform polarization compensation around a certain axis on the Poincaré sphere. Whereas the use of only one reference signal restricts the polarization control to one axis, any output SOP can me mapped into the original input one by employing a pair of non-orthogonal reference signals. We show the effectiveness of this scheme, by measuring the power penalty, which may be interpreted as the optical quantum bit error rate ($QBER_{opt}$) in this context [3], that a polarization sensitive detector would have under polarization control. This is the first time a complete polarization control is performed by using the intensities of the reference signals as feedback signals to the polarization controller.

Our scheme is well suitable to quantum communications systems, such as polarization-coded Quantum Key Distribution via optical fibers. Since it does not use the main signal as feedback, full real-time polarization control can be achieved, thus improving previously suggested schemes for this purpose [4].

## 2. Theory

Full control of any polarization state can be achieved if two non-orthogonal polarization states, henceforth called $S_1$ and $S_3$, are sent across the fiber as reference states. One can choose, for instance, without loss of generality, the horizontal and +45° linear polarization states, respectively, which are canonically conjugate non-orthogonal states. At the end of the fiber we place a polarization controller which performs a series of two rotations, $R_1$ and $R_3$, bringing the output SOP into the desired state. Calling T the Jones matrix representing the transformation over the fiber, our control system requires that the following system of equations holds:

$$\begin{cases} R_1 T S_1 = S_1 \\ R_3 S_1 = S_1 \\ R_3 R_1 T S_3 = S_3 \end{cases} \quad (1)$$

That is, the role of the unitary transformations $R_1$ and $R_3$ is to cancel out the effects due to birefringence in the fiber for the input SOPs $S_1$ and $S_3$, respectively. Note that $R_3$ does not have any effect upon $S_1$, which means that it is a rotation around the axis, in the Poincaré sphere defined by $S_1$ and its orthogonal state. The first two lines of Eq. (1) can be rewritten in matrix form as follows:

$$R_1 = \begin{pmatrix} 1 & 0 \\ 0 & e^{j\phi} \end{pmatrix} T^{-1}, \quad R_3 = \begin{pmatrix} 1 & 0 \\ 0 & e^{j\theta} \end{pmatrix} \quad (2)$$

Where $\phi$ and $\theta$ can take any value between 0 and $2\pi$. Combining this result with the third line of Eq. (1), we get:

$$\begin{pmatrix} 1 & 0 \\ 0 & e^{j(\theta+\phi)} \end{pmatrix} \begin{pmatrix} 1 \\ 1 \end{pmatrix} = \begin{pmatrix} 1 \\ 1 \end{pmatrix} \quad (3)$$

Eq. (3) holds if and only if $\theta + \phi = 0 \mod 2\pi$. This means that $R_3 R_2 T$ must be equal to the identity, or equivalently, $R_3 R_1 = T^{-1}$, which clearly shows that the effect of the polarization controller is to nullify the polarization rotations generated by $T$. Any polarization state passing through the fiber and the two rotators $R_1$ and $R_3$ will be preserved. Of course, signal and references have all the same optical frequency and the control system must distinguish between them to perform the correct rotations at each controller. This is easily done either by time multiplexing references and signal or by dithering the signal and references at different frequencies and filtering at the receiver.

Although this method can precisely control all polarization states sent throughout the fiber, co-propagation of reference signals along with a quantum channel is very problematic. References extinction ratio must be extremely high to allow photon counting in the quantum channel without too many false counts. An alternative approach is to send the reference signals at different wavelengths, in co-propagating side channels. These channels could also be simultaneously used to synchronize the transmitter and receiver and send disclosed information between them. Here also the filtering needs are high, but the quality of nowadays Dense Wavelength Division Multiplexing (DWDM) filters is good enough to insure the needed isolation. Synchronization with side channel transmission in quantum key distribution was indeed achieved in recent experiments [5]. Clearly, the fluctuating birefringence, which is the matter of our control system, is also responsible by differences between the signal channel, which we will now call $S_2$ and the two reference channels $S_1$ and $S_3$. However, we will show that if the PMD of the fiber is not too high, good quality control can be achieved.

Now we require that the polarization $S_2$ of a signal at wavelength $\omega_0$ to be controlled by employing reference signals at $S_1$ and $S_3$ at different wavelengths, given by $\omega_1 = \omega_0 - \Delta\omega$ and $\omega_3 = \omega_0 + \Delta\omega$. It is also important to stress the dependence of $T$ on the wavelength, thus we write $T = T(\omega)$. Hence, we get a new system of equations, similarly to Eq. (1):

$$\begin{cases} R_1 T(\omega_1) S_1 = S_1 \\ R_3 S_1 = S_1 \\ R_3 R_1 T(\omega_3) S_3 = S_3 \end{cases} \quad (4)$$

Now we proceed exactly as done in the monochromatic case. Noticing that $T(\omega_3) = T(\omega_1) + 2\Delta\omega (\partial T/\partial\omega)$ we get:

$$\begin{pmatrix} 1 & 0 \\ 0 & e^{j(\theta+\phi)} \end{pmatrix} \left( I + 2\Delta\omega T^{-1} \frac{\partial T}{\partial \omega} \right) \begin{pmatrix} 1 \\ 1 \end{pmatrix} = \begin{pmatrix} 1 \\ 1 \end{pmatrix} \quad (5)$$

Where both $T^{-1}$ and $\partial T/\partial\omega$ are calculated at $\omega_1$. The same solution found for the monochromatic case, i.e. $\theta + \phi = 0 \mod 2\pi$, is still valid if the condition below holds:

$$2\Delta\omega \left\| T^{-1} \frac{\partial T}{\partial \omega} \right\| << 1 \quad (6)$$

As the matrix norm in Eq. (6) is precisely half the DGD value $\tau/2$ [6], the condition can be restated as $\tau \Delta\omega << 1$.

## 3. Experiment

In our experiment we used the optical frequency domain to separate the reference signals from the signal channel. The experimental set-up is presented in Fig. 1. References and signal were separated by 0.8 nm and launched into the fiber (8.5 km long, 0.54 ps PMD) via a WDM. A four-plate piezoelectric polarization controller, followed by a second controller, performed the rotations $R_1$ and $R_3$, according to the previous description. The control signals were launched from Bob' side, counter propagating the quantum channel, thus limiting the interference of the side channels to the Rayleigh backscattered light. A second filter was used to further reduce the crosstalk to below the dark count level of the single photon counting module in channel 2. At Alice's side of the link, after passing through all polarization controllers, the control channels were dropped through port 3 of an optical circulator and separated by a second WDM. Channels 1 and 3 were detected (photodiodes $PD_1$ and $PD_3$) after passing through the corresponding linear polarizers ($P_1$ and $P_3$) with polarization controllers adjusting the axes of $R_1$, $R_3$ and the polarizers $P_1$ and $P_3$ such that once the polarization is adjusted in channel 1 it is invariant to rotations performed by $R_3$. Efficient control was obtained with received powers as low as -20 dBm in the control channels. At Alice's side, the quantum channel was launched after a polarization controller and a variable attenuator, so that polarization measurements could be made either with classical light or with single photon counting measurements by placing a polarimeter at Bob's end or a polarization controller followed by a linear polarizer and a gated photon counter. The rejection ratio of the polarizers was better than 40 dB. Photon counting measurements were made with 0.2 photons per 2.5 ns gate pulse at 100 kHz rate.

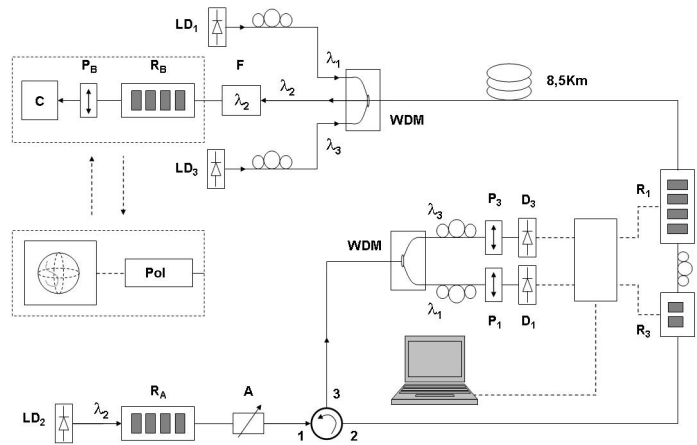

Fig. 1. Experimental set-up: The three loops represent manual polarization controllers, R: electrically driven polarization controllers, P: linear polarizers, D: classical photodetectors, C: single photon counting module, LD: Laser diodes, A: attenuator and Pol: polarimeter.

Polarization controllers (40 kHz bandwidth) and fast A/D conversion (500 MHz) allow efficient control under fast variations in the transmission line. Figure 2 presents the detected intensity through a polarizer in channel 2, after a fast polarization rotation was induced in the transmission link. The system recovers very fast to 90% received power (~2.5 ms), and full signal recovery is obtained in times smaller than 10 ms. This means that the technique can be used to control polarization states even in fast varying transmission lines.

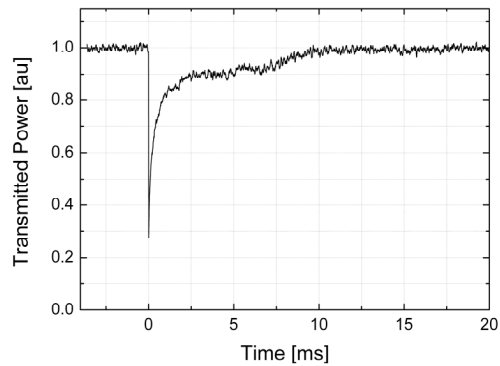

Fig. 2. Signal recovery in channel 2 controlled by channels 1 and 3.

Figure 3 displays the evolution on the Poincaré sphere of the output polarization of channel 2 with and without polarization control, over 2 hours during which the PMD of the fiber was forced to evolve by changing its temperature. Clearly, a polarization based quantum channel would never perform in such an uncontrolled fiber, and that's the reason why polarization encoding is so difficult to be implemented in optical fibers. When the polarization control is turned on, the stabilization of two non-orthogonal polarization states is clear.

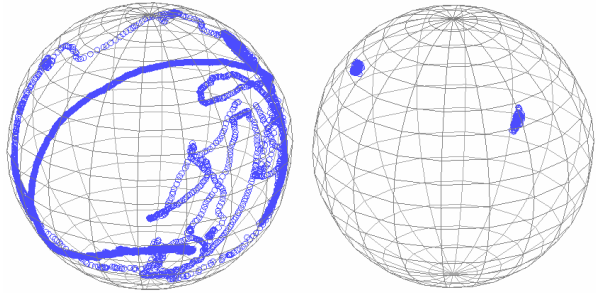

Fig. 3. Left: uncontrolled time evolution of a single output SOP. Right: evolution two output non-orthogonal SOPs with full polarization control

Considering the figures used in the experiment, Eq. 6 gives $\tau \Delta \omega \sim 0.3$, a value not so small if compared to the unity. However, the performance of the experiment is striking, which means that Eq. 6 seems too restrictive. In fact, it was deduced considering a worst case scenario, as the solution $\theta + \phi = 0 \pmod{2\pi}$ is not necessarily the unique solution for Eq. 5. Hence, it is possible that the control algorithm finds better solutions in the general case, somehow relaxing the condition stated in Eq. 6.

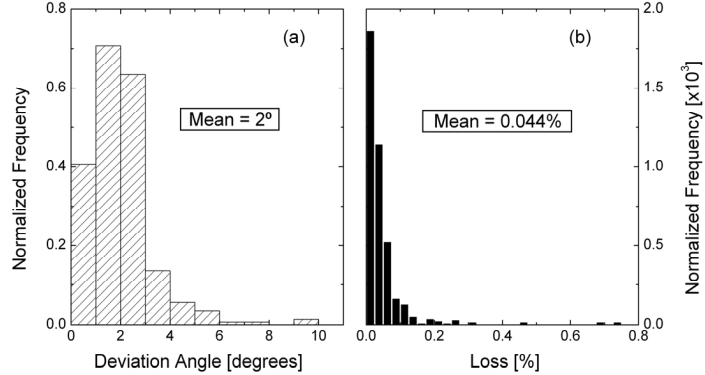

Fig. 4. (a) Statistics of the deviation angle between the received and target SOP in the Poincaré sphere. (b) Corresponding added power penalties under full polarization control.

A better evaluation of the efficiency of the control scheme can be seen in figure 4(a), which presents the statistical distribution of the deviation angle (in the Poincaré sphere) between the actual polarization and the target polarization. We observe that all deviation angles are smaller than $10^o$ with a mean value of only $2^o$. This means that the worst case added channel power loss due to the polarization control in this experiment was smaller than $\cos^2(5^o) \sim 0.8\%$. Fig. 4(b) displays the statistical distribution of the power loss that would be added by the polarization control system due to the mismatch between the controlled and target SOPs during the experiment. Of course, this distribution depends on the temporal evolution of the optical fiber transfer function and can be degraded in case of severe fiber vibrations. Moreover, if the PMD of the fiber is much higher than 0.5 ps the accuracy of the control will decrease, and the SOP of the qubits will swing around its target SOP, thus widening the distribution of deviation angles. However, considering the recovery time of the system and the fact that the channel loss is given by the squared cosine of half the deviation angle in the Poincaré sphere, the precision of our polarization control system leaves a margin good enough to override the sensitivity of most quantum communications detection systems and is fully compatible with accepted optical QBER in quantum communications, which is usually around 1%.

Now we must show that the proposed system is able to operate effectively in the single-photon counting regime. One could state, for instance, that detrimental influence of the strong reference beams, such as cross-talk of the Rayleigh-scattered light from the control channels in the fiber or due to insufficient isolation of the DWDMs or filters, could create an overwhelmingly high number of noise counts such that the QBER would exceed the maximum acceptable value. In order to evaluate the isolation between the quantum and control channels, we increased the launched power of the side channels to +5 dBm each, much higher than the minimum needed for an efficient control. In this condition we still observed that the crosstalk was smaller than the dark counts of our InGaAs single photon counter, which has a dark count probability of $4 \times 10^{-5}$ per nanosecond. Fig. 5 shows the single photon counts at Bob for two fixed polarization states (linear 0º and 45º) at Alice, corresponding to the same bit value in the two non-orthogonal bases of a BB84 protocol for instance. In each case, Bob uses a polarization controller to change the polarizer angle along the equator of the Poincaré sphere. It can be seen that, when Bob's polarizer is orthogonal to the state sent by Alice, the total noise counts are entirely comprised of the single photon counter module's dark counts. This means that all other sources of noise are negligibly small, which demonstrates the feasibility of the proposed setup.

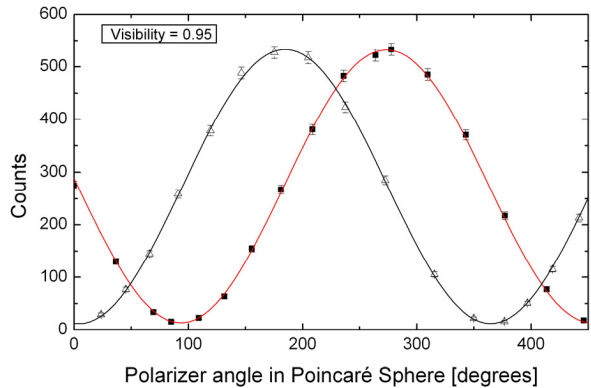

Fig. 5. Single photon counts at Bob for two fixed polarization states (linear 0º and 45º) sent by Alice controlling the polarization through 8.5 km single mode fiber.

## 4. Conclusion

A polarization control system using two wavelength-multiplexed non-orthogonal reference signals has been theoretically and experimentally evaluated. It has been shown that if references and signal are separated by 0.8 nm via a WDM, a complete control of multiple polarization states can be achieved, with an average added signal power loss of only 0.044% for arbitrary input and output pairs of polarization states. This result shows that quantum communications using polarization encoding can be made possible using an optical fiber as the quantum channel, as the average added optical QBER is in orders of magnitude lower than the detection QBER introduced by the single photon counting module.

## Acknowledgment

The authors are grateful to N. Gisin for lending the single photon counting module used in the experiments.